# Real-time motion amplification on mobile devices

Henning U. Voss

Cornell University, Cornell MRI Facility, Ithaca, NY, USA

Weill Cornell Medical College, Dept. of Radiology, New York, NY, USA

Email: hv28@cornell.edu

ORCID ID: 0000-0003-2811-2074

**Abstract**
A simple motion amplification algorithm suitable for real-time applications on mobile devices, including smartphones, is presented. It is based on motion enhancement by moving average differencing (MEMAD), a temporal high-pass filter for video streams. MEMAD can amplify small moving objects or subtle motion in larger objects. It is computationally sufficiently simple to be implemented in real time on smartphones. In the specific implementation as an Android phone app, MEMAD is demonstrated on examples chosen such as to motivate applications in the engineering, biological, and medical sciences.

# 1 Introduction

Motion amplification is a fascinating tool that can make unperceivable motion visible by computational amplification. Many motion amplification algorithms are based on principles that have been developed decades ago [1], but really gained traction more recently with the improvement of computational power. This development also enabled more accurate and powerful motion amplification algorithms including motion microscopy [2-6]. For a review of motion amplification algorithms, see Le-Ngo and Phan [7]. Present applications of motion amplification include motion within the brain [8], the monitoring of vital signs [9], and the detection of subtle facial expressions [10-12].

It would be interesting to evaluate whether motion amplification is possible in real time on mobile devices such as smartphones. Real-time applications of motion amplification, i.e., with instantaneous results and without any processing on other equipment than the smartphone itself, could be beneficial whenever an instant exploration of motion is desired. For example, in remote biological field studies, for pulse measurement in communication apps, or in on-site diagnostics of machinery. To approach this goal, this research took a step back from the previously mentioned state-of-the-art algorithms, thereby developing a computationally simple motion amplification algorithm that is less demanding on processing hardware and thus suitable for real-time smartphone apps.

The organization of this manuscript is as follows: The Methods section describes the MEMAD motion amplification model developed here, its implementation on smartphones, and a simple example. In the Applications section, three real-time smartphone applications are shown: Visualizing vibrations of an engine, estimating the pulse of a person from live video, and real-time video trapping of small animals. In addition, the MEMAD model is being applied to a set of benchmark videos that have been published with other motion amplification models before. A discussion mentioning limitations of this real-time smartphone solution concludes this article. The video files are an important part of the manuscript and can be viewed or downloaded at https://doi.org/10.6084/m9.figshare.20084981.v2.

# 2 Methods

### The MEMAD Model

The motion enhancement by moving average differencing (MEMAD) model used here is specified as follows. The camera stream is described by RGB images of size $M \times N$ at sample time t. The MEMAD filter maps each pixel intensity value $I_k$ (k = R, G, B) at position (x = 1...M, y = 1...N) via

$$I_k(x,y,t) \rightarrow \beta I_k(x,y,t) + \alpha \left( I_k(x,y,t) - \frac{1}{m} \sum_{i=1}^{m} I_k(x,y,t-i) \right). \qquad (1)$$

At initialization, pixel intensity values that are undefined are set to zero. This filter compares the present intensity values to the moving average of m immediate prior values, and is thus a high-pass filter sensitive to changes in intensity values. The high-pass values are amplified by a factor α and then added to the present intensity, which can be dimmed by a factor β in order to make the amplified differences more stand out against a darker background image. This filter has to be understood modulo the number representation of pixel intensities in specific hardware applications; in the present application, RGB values are represented by 8-bit unsigned integers and thus have a range of 0 to 255. In all applications described here, mapped intensities exceeding this range are set to 0 or 255, for the case of underflow and overflow, respectively.

The motivation for the model Eq. (1) is computational simplicity, allowing for real-time computation on contemporary smartphones. More advanced motion detection algorithms have been proven difficult to implement on smartphones in real-time due to their higher computational demand.

**Implementation Details**

Figure 1 shows the SIMULINK flow diagram of the MEMAD model for Android mobile device implementations. The file itself is provided as a data file (see section Description of data) but the following description should contain all information needed to reproduce the application, too. The camera stream from the Android device consists of three separate channels for RGB encoding. These RGB streams are each delayed by up to m = 5 sampling time steps and then fed into the MATLAB function MEMAD, which performs the actual filtering given by Eq. (1). The filter result is then streamed to the device's display in real time.

In all applications shown, the parameters of Eq. (1) were m = 5, $\alpha$ = 16 or 32, $\beta$ = 0.5 or 1.0, as specified.

*Camera block*

The resolution of the Android camera block was set to 240 × 320 pixels, and its sampling rate to 30 fps. Depending on the chosen camera resolution, the MEMAD function might not be fast enough to process the images in real time within this sampling time. Therefore, it was taken advantage of the SIMULINK feature that sampling rates can be backwards adjusted. In other words, the output rate on the device's display can appear to be lower than 30 fps. This was observed in the present applications with the given image matrix size.

*Delay blocks*

At the beginning of the stream, the initial values of the delay blocks are set to zero, and the input processing of the delay blocks is "Elements as channels" to define sample-based processing.

*MATLAB function block*

The MEMAD block content is

```
function [Rout,Gout,Bout] = MEMAD(R,G,B,Rbuff1,Gbuff1,Bbuff1,Rbuff2,Gbuff2,Bbuff2, Rbuff3,Gbuff3,Bbuff3,
   Rbuff4,Gbuff4,Bbuff4, Rbuff5,Gbuff5,Bbuff5)

alpha=16; beta=1;

Rout = beta*R+ alpha*( R-Rbuff1/5-Rbuff2/5-Rbuff3/5-Rbuff4/5-Rbuff5/5 );
Gout = beta*G+ alpha*( G-Gbuff1/5-Gbuff2/5-Gbuff3/5-Gbuff4/5-Gbuff5/5 );
Bout = beta*B+ alpha*( B-Bbuff1/5-Bbuff2/5-Bbuff3/5-Bbuff4/5-Bbuff5/5 );
```

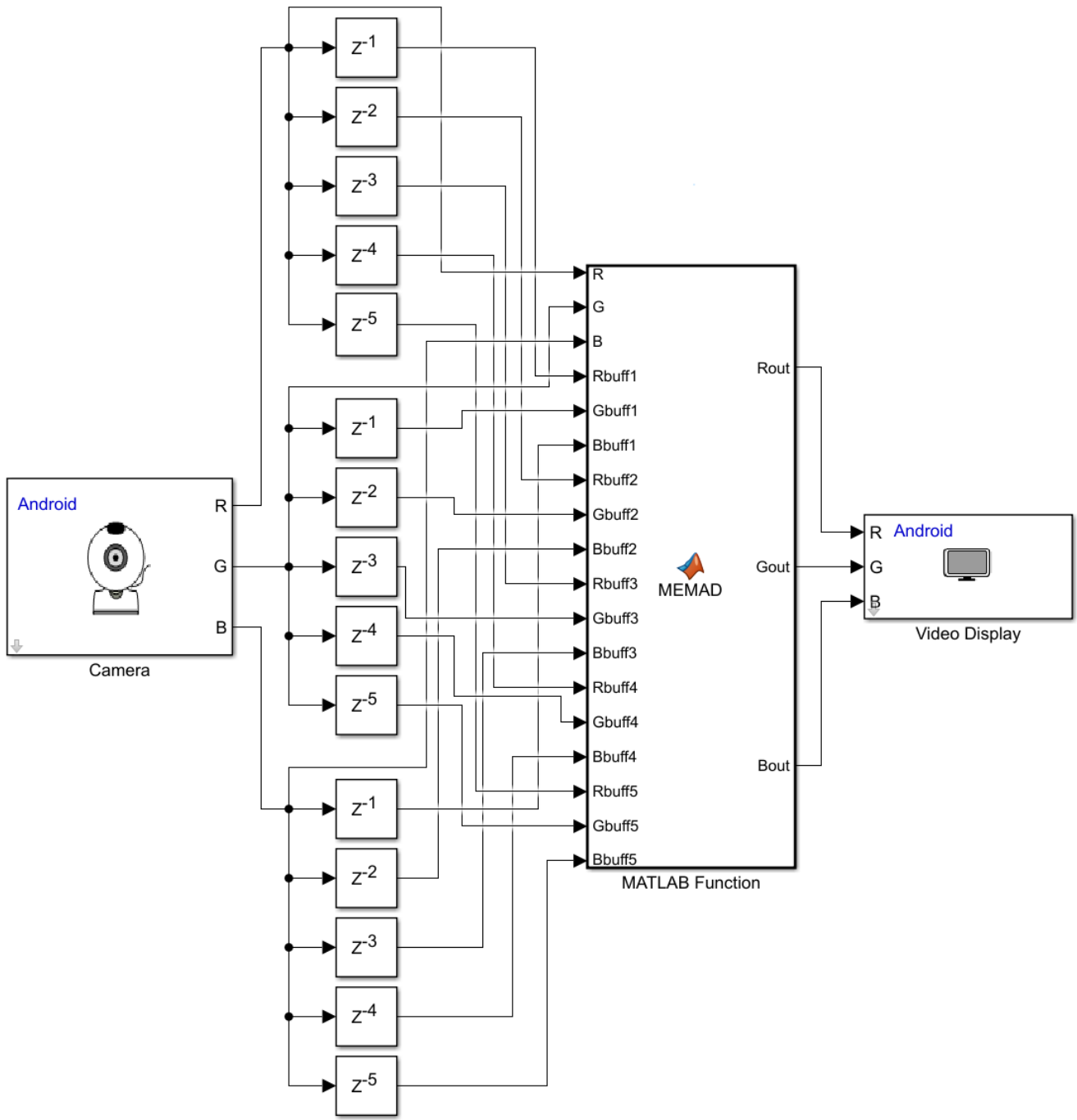

**Figure 1. SIMULINK model for real-time motion amplification on a smartphone.** The central element is the MEMAD video filter that receives time-delayed RGB frames from the phone camera and passes them to the phone's screen in real time.

*Hardware and software versions*

The SIMULINK version was 10.2 (R2020b). SIMULINK was used together with Android Studio Chipmunk 2021.2. with Android SDK Build-Tools 33-rc4 and Android SDK Platform 30 for Android 11.0 (R).

The device used for testing the filter was a Motorola Moto G Stylus, hardware version pvt, RAM size 4 GB,

ROM size 128 GB, display size 2300 × 1080, front camera 16 MP, and rear camera 48 MP. The software version was Android 11, kernel version 4.14.180-perf+ #1 Sun Mar 27 09:24:05 EDT 2022, build number RPRS31.Q1-56-9-15.

### Recording

For recording MEMAD videos the smartphone was mounted on a tripod. The phone's screen was captured in real time with the XRecorder app (InShot Inc., update 4/28/2022). Unfiltered videos for comparison were recorded with the phone's in-built camera app. Recorded files were transferred to a PC for inspection and selection of sections for the video files.

### Example – Falling Rain

The MEMAD algorithm was tested in real time on an Android smartphone. The purpose was to amplify small moving objects with low background contrast. Fast moving objects of the size of one to just a few image pixels often do not stand out clearly in videos, in particular if their background contrast is low. The amplification of subtle motion in larger objects will be shown in the Applications section below. Video file rain_a.mp4 (see data files) is an unfiltered capture of a light rainfall. The motion amplified MEMAD results are shown in video files rain_b.mp4 and rain_c.mp4. Video file rain_b.mp4 has amplification α = 32 and dimming factor β = 1, and was recorded right after rain_a.mp4. Video file rain_c.mp4 has the same amplification but β = 0.5. As expected, falling raindrops are amplified in real time by the MEMAD model Eq. (1), and the case of β < 1 relatively amplifies movement stronger at the cost of still image component quality. Snapshots of the videos are shown in Fig. 2.

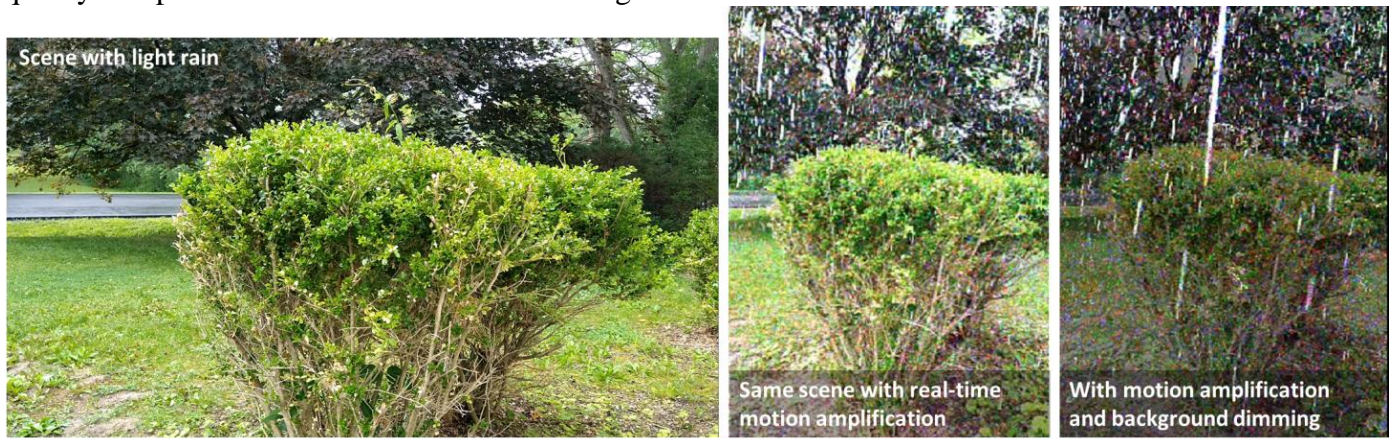

**Figure 2: Example – Light rain.** The left panel shows a snapshot of the video file rain_a.mp4, recorded with the built-in Android camera app. The middle panel shows the same scene with the real-time MEMAD motion amplification app (amplification α = 32 and dimming factor β = 1; video file rain_b.mp4). The right panel has the same amplification but β = 0.5; video file rain_c.mp4.

## 3  Applications

An overview of real-time applications is provided in Table 1.

**Engineering Application - Engine Vibrations**

Vibrations of engines and other machinery are well-known targets for industrial motion amplification applications. A video of the amplified vibrations of an engine in a passenger car was recorded in real time on an Android device using MEMAD. Video file engine_a.mp4 contains an unfiltered high-resolution video obtained from the phone's camera app for comparison. Video file engine_b.mp4 contains a screen captured MEMAD video of the engine with amplification $\alpha = 32$ and dimming factor $\beta = 1$. It reveals apparent engine vibrations, whereas the surrounding parts of the car are mostly still. These vibrations appear more enhanced with a dimming factor of $\beta = 0.5$ in video file engine_c.mp4.

Of note, in this perceptual approach, no motion trajectories are estimated, and the perceived motion results from the way the brain interprets intensity and color change. This will be detailed in the Discussion section.

**Telehealth Application - Remote Pulse Sensing**

It has been demonstrated before that facial skin color can change in synchrony with the heartbeat, when amplified under laboratory conditions [13, 14].

In order to demonstrate that a mobile device with low resolution and low sampling rate can provide imaging data allowing for the real-time estimation of the heart rate of a person, a video of the author was recorded in real-time on an Android device using MEMAD. The parameters were $\alpha = 16$ and $\beta = 1$. The real-time MEMAD video pulse.mp4 turned out to be grainy and does not allow for an instant appreciation of the effects of heartrate on skin color. However, the heart rate could be determined accurately on the phone from the MEMAD video by running a script on the MATLAB Mobile app. The script performed a power spectral analysis of the average image intensity of the video, as shown in Figure 3.

The video file pulse.mp4 also shows apparent head motion, an effect similar to the vibration of the engine in file engine_b.mp4, which is being discussed in the Discussion section.

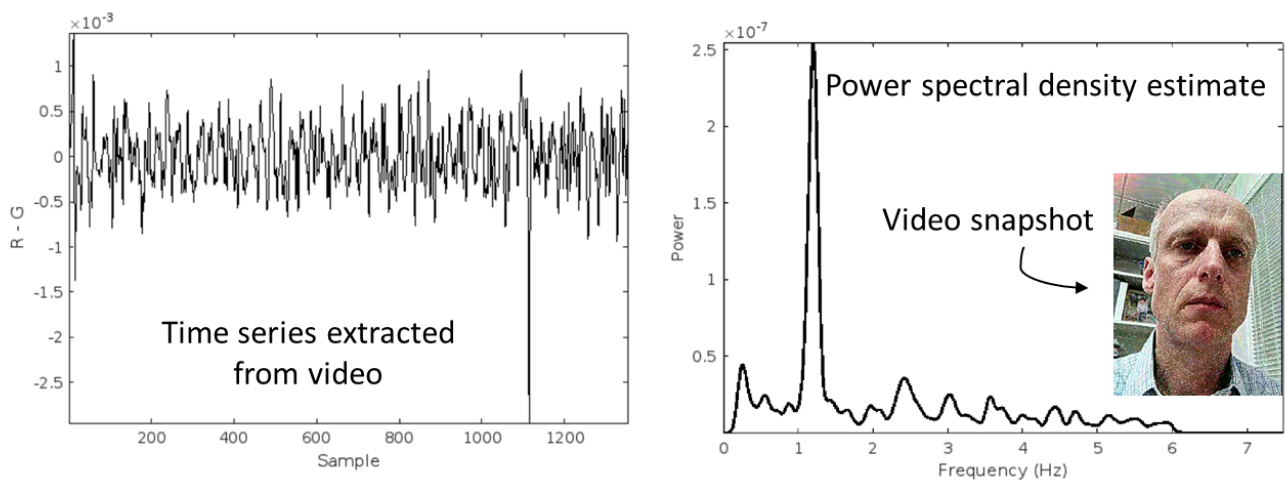

**Figure 3: Pulse from video stream.** The left panel shows a screenshot of the difference between the red and green RGB channels of a video stream on an Android smartphone, amplified in real-time by MEMAD. The video stream was spatially averaged over the whole frames, and the resulting single time series band-pass filtered between 0.2 and 6 Hz. These computations were performed on the phone using MATLAB Mobile. The right panel shows a screenshot of a power spectral analysis of the time series, again obtained with MATLAB Mobile. It has a peak power at a frequency of 1.2 Hz, corresponding to an average pulse of 72 bpm. Palpation of the pulse on the wrist for 1 min immediately before the experiment yielded an average pulse of 73 bpm.

**Biological Application – Wildlife Camera Trapping of Small Animals**

Wildlife camera trapping is a common tool used in biological research and other applications for imaging animals with a seclusive or nightly lifestyle. However, small animals such as insects, spiders, etc. often live in plain sight yet remain undetected by us due to their small size, camouflage, or generally low background contrast.

In order to test the hypothesis that small animals with low contrast can be revealed by means of their movements, a motion-enhanced video of a nature scene was recorded in real time on an Android device. The video file smallanimals_a.mp4 is a high-resolution video of the nature scene. The occasional flyby of an insect and a bug on a stone can be seen, but otherwise the video makes a quite tranquil impression. Video file smallanimals_b shows the same scene motion-enhanced in real-time with MEMAD, with amplification $\alpha = 16$ and dimming factor $\beta = 0.5$. The latter parameter had been chosen to make moving objects stand out more relative to the background. It was recorded with the same device back-to-back to the previous video and reveals a busy scene filled with small flying animals. It is, however, sometimes difficult to discern which variability is due to moving animals and which is due to amplified noise.

**Offline Applications**

In order to facilitate comparison with published offline motion amplification algorithms, MEMAD was applied offline to a series of video files published by the CSAIL lab at MIT [6]. These videos, along with motion amplified versions, can be downloaded from http://people.csail.mit.edu/nwadhwa/phase-video/. Here, sometimes shortened versions of the videos were used for processing.

In offline applications, there is more freedom in choosing the MEMAD parameters, and the model Eq. (1) was expanded by inclusion of an additional parameter n to define a bandpass filter. The offline model is

$$I_k(x,y,t) \rightarrow \beta I_k(x,y,t) + \alpha \left( \frac{1}{n} \sum_{i=1}^{n} I_k(x,y,t-i+1) - \frac{1}{m} \sum_{i=1}^{m} I_k(x,y,t-i) \right). \tag{2}$$

For n = 1, this model coincides with the previous model. In offline applications, it is further possible to increase the parameter m to larger values than 5.

The video files MIT_baby.mp4, MIT_eye.mp4, MIT_face.mp4, MIT_woman.mp4, MIT_engine.mp4, and MIT_crane.mp4 contain the unfiltered MIT video side-by-side to the motion amplified video. Values for the used parameters α, n, and m are provided in Table 2.

## 4  Discussion and Conclusion

The goal of this work was to demonstrate that basic motion amplification can be executed in real time on a mobile device. Towards this end, a very simple motion amplification algorithm has been developed – Motion enhancement by moving average differencing (MEMAD). The functionality of MEMAD has been demonstrated in four examples; the real-time enhancement of falling rain drops, real-time amplification of engine vibrations, real-time image capturing for pulse measurement, and real-time small animal video trapping. In the engine and pulse examples, MEMAD generated a motion illusion from subtle movements; in the rain and small animal examples, small moving objects with low background contrast were made visible.

**Motion Illusion**

Whereas the intensity amplification of small moving objects (as in the rain and small animal tracking examples) follows directly from the high-pass properties of the filter, the source of global motion perception (as in the vibrating engine and apparent head movements in the pulse video) seems not to be so trivial. As has been reviewed by Refs. [1, 15], motion illusions like these can arise from local phase changes that are perceived as global motions. In the engine example, the edge intensities are varying the most and are amplified relatively more by the MEMAD filter than the more homogeneous parts of the engine. A possible explanation could be that the amplified edge intensity variations are approximating local phase changes, which then provide the global motion illusion [1].

**Use of SIMULINK**

The specific route to implementation on Android devices involved the use of SIMULINK. This route was chosen in order to make the communication between the MEMAD function and the Android camera and screen as simple as possible, by utilizing existing Android blocks in SIMULINK and MATLAB matrix processing functionality. In addition, SIMULINK's sampling and buffering management proved to be very helpful in this implementation. This application could also have been implemented without the use of MATLAB and SIMULINK, as has been done before by Chambino [14] for other real-time motion amplification algorithms on a PC with Android software. Such a direct implementation could result in enhanced processing speed.

**Possible Clinical Application**

An ongoing research topic with potential clinical implications is the detection of subtle cardiac pulsatility in the human brain [8, 16-18]. High-pass filters such as MEMAD might have potential to enhance this kind of motion in temporal sequences of magnetic resonance images. In addition, real-time mobile motion amplification apps could add to the emerging field of augmented reality-based apps for biomedical imaging data on mobile devices [19].

**Limitations**

It stands to reason that a sampling rate of only 30 fps or less can only resolve oscillations up to the corresponding Nyquist frequency, and that this smartphone application cannot compete with advanced offline motion amplification systems currently used in research labs and industry. This might not be a problem for the raindrop, pulse, and small animal trapping examples but could be for the engine, and the amplified engine videos might be affected by finite sampling effects, such as aliasing.

In all real-time applications shown here, a fixed filter order of 5 has been used as a compromise between performance and computational demands. Depending on the specific application, a higher filter order for slow motion / lower filter order for fast motion might be better suited to amplify motion.

**Conclusion**

It has been demonstrated that perceptual motion amplification can be executed in real time on a mobile device with limited computational power, exemplified on an Android smartphone. In order to realize this, first, a simplified perceptual motion amplification algorithm has been devised and then applied to several examples, which demonstrated both motion illusion from subtle movements and from small moving objects with low background contrast. Limitations compared to previously published motion amplification approaches , which are not meant to work in real time yet, were discussed.

# 5   Description of Data Files

The data files can be viewed or downloaded at https://doi.org/10.6084/m9.figshare.20084981.v2.

**Android App Deployment File**

The SIMULINK file MoveIt7_16_1_240320.slx contains the application. In order to deploy it to an Android device as the app MoveIt7_16_1_240320, it is necessary to open it with SIMULINK and to follow the instructions regarding the installation of additional software such as Android Studio and MATLAB support packages for Android. Using existing equivalent SIMULINK support packages for iOS devices, it should be straightforward to deploy the app to iOS devices, too, but this was not tested here. The default parameters are $\alpha = 16$, $\beta = 1.0$, and the camera resolution is $240 \times 320$ pixels. The former two parameters can easily be changed by the user in the MEMAD module, which is a MATLAB function that can be edited. The camera resolution options depend on the connected device, and, therefore, the desired resolution needs to be set manually after refreshing device options and before deployment to the device. The Methods section details tested SIMULINK, MATLAB, and Android Studio software versions.

**Video Files**

The real-time video files are provided in Table 1. The offline video files are provided in Table 2.

| Section II.D: Example – Falling rain | | |
|---|---|---|
| 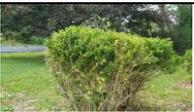 | **rain_a.mp4** | High-resolution video of a rainy scene acquired with the smartphone's camera app |
| 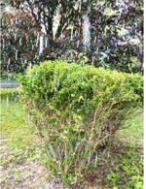 | **rain_b.mp4** | MEMAD motion amplified video with α = 32, β = 1 |
| 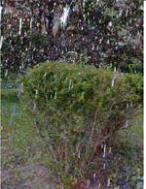 | **rain_c.mp4** | MEMAD motion amplified video with α = 32, β = 0.5 |
| Section III.A: Engineering application - Engine vibrations | | |
| 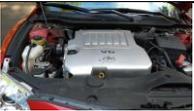 | **engine_a.mp4** | High-resolution video of a car engine acquired with the smartphone's camera app |
| 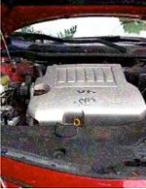 | **engine_b.mp4** | MEMAD motion amplified video with α = 32, β = 1 |
| 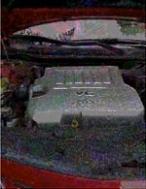 | **engine_c.mp4** | MEMAD motion amplified video with α = 32, β = 0.5 |
| Section III.B: Telehealth application - Remote pulse sensing | | |
| 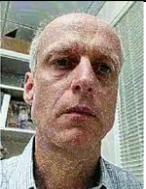 | **pulse.mp4** | MEMAD video stream from the smartphone's front camera, α = 16, β = 1. |
| Section III.C: Biological application – Wildlife camera trapping of small animals | | |
| 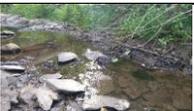 | **smallanimals_a.mp4** | High-resolution video of a nature scene acquired with the smartphone's camera app |
| 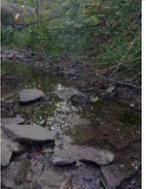 | **smallanimals_b.mp4** | MEMAD motion amplified video with α = 16, β = 0.5 |

**Table 1. Real-time video files**. The parameters refer to Eq. (1). In all examples, m = 5.

| | | |
|---|---|---|
| 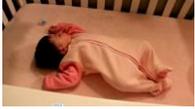 | **MIT_baby.mp4** | MIT video of a sleeping baby and offline MEMAD motion amplified video with n = 5, m = 30, α = 4. Video sampling rate = 30 fps. |
| 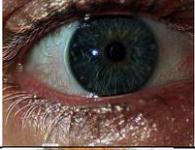 | **MIT_eye.mp4** | MIT video of an eye and offline MEMAD motion amplified video with n = 8, m = 30, α = 4. Video sampling rate = 30 fps. |
| 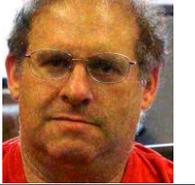 | **MIT_face.mp4** | MIT video of a face and offline MEMAD motion amplified video with n = 8, m = 30, α = 16. Video sampling rate = 30 fps. |
| 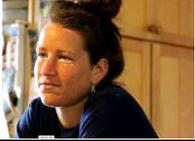 | **MIT_woman.mp4** | MIT video of a person and offline MEMAD motion amplified video with n = 15, m = 60, α = 8. Video sampling rate = 60 fps. |
| 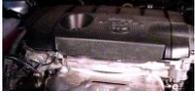 | **MIT_engine.mp4** | MIT video of a car engine and offline MEMAD motion amplified video with n = 6, m = 25, α = 32. Video sampling rate = 25 fps. |
| 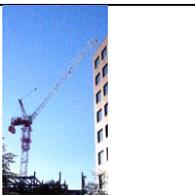 | **MIT_crane.mp4** | MIT video of a slightly swinging crane and offline MEMAD motion amplified video with n = 6, m = 24, α = 32. Video sampling rate = 24 fps. |

**Table 2. Offline video files.** The parameters refer to Eq. (2). In all examples, β = 1.